\documentclass[aps,amsmath,twocolumn,amssymb,titlepage,10pt]{revtex4-1}
\usepackage[T1]{fontenc}
\usepackage[utf8]{inputenc}
\usepackage{amsmath}
\usepackage{braket}
\usepackage{amsfonts}
\usepackage{comment}
\usepackage{graphicx}
\usepackage{hyperref}
\usepackage[capitalize]{cleveref}
\usepackage{amssymb}
\usepackage{amsmath}
\usepackage{mathtools}
\usepackage{todonotes}

\begin{document}
\title{Quantifying  nonuniversal corner free-energy contributions in \\ weakly-anisotropic two-dimensional critical systems}
\author{Florian Kischel}
\author{Stefan Wessel}
\affiliation{Institute for Theoretical Solid State Physics, RWTH Aachen University, Otto-Blumenthal Str.~26,  52074 Aachen, Germany}
\begin{abstract}
We derive an exact formula for the corner free-energy contribution of weakly-anisotropic two-dimensional critical systems in the Ising universality class on rectangular domains, expressed in terms of quantities that specify the  anisotropic  fluctuations. The resulting expression agrees with numerical exact calculations that we perform for the anisotropic triangular Ising model and quantifies the nonuniversality of the  corner term for anisotropic critical two-dimensional systems.  Our generic formula is expected to  apply also to other weakly-anisotropic critical two-dimensional systems that allow for a conformal field theory description in the isotropic limit. We consider the 3-states and 4-states Potts models as further specific examples.
\end{abstract}
\maketitle

\section{Introduction}\label{Sec:Introduction}
Critical phenomena are of fundamental importance to  modern condensed matter physics. Of particular interest are physical quantities with a universal quality at criticality, i.e., those that take on specific values characteristic of the underlying universality class (UC), such as critical exponents. For two-dimensional (2D) systems in particular, a large amount of analytic results is available in this respect from both exact solutions of specific model systems as well as based on the general framework of conformal field theory (CFT)~\cite{Cardy1988a,DiFrancesco1997}. A prominent example, which is central to the present study, is the prediction by Cardy and Peschel~\cite{Cardy1988} of a logarithmic  contribution to the free energy from corners along the boundary of a confined conformal invariant bulk system, and which is proportional to the central charge in the CFT limit. 

More specifically, for a 2D critical system, such as the Ising model at its critical temperature $T_c$, the free energy density $f^c$ (in units of the thermal energy $k_B T_c$) scales for large systems with free (open) boundaries, area size $A$ and edge length $E$ as~\cite{Wu2012,Wu2013,Izmailian2017}
\begin{equation}\label{Eq:fobc}
f^c=f^c_{\rm bulk} + f^c_{\rm surface} \frac{1}{E} + f^c_\mathrm{corner} \frac{\ln A}{A} + f^c_2 \frac{1}{A} +\dots,
\end{equation}
where in addition to the bulk ($f^c_{\rm bulk}$) and surface ($f^c_{\rm surface}$) contributions the corner term $f^c_\mathrm{corner}$ appears in case that the boundary of the spatial domain contains  corners separated by otherwise straight edges. The case of a rectangle is of particular importance, and we 
mainly focus on such domains.
In Eq.~(\ref{Eq:fobc}), the expansion to higher order contributions has been terminated. Indeed, similarly to the bulk and surface contribution, the higher order terms depend on microscopic details, whereas the value of $f^c _\mathrm{corner}$, while geometry-dependent, has been derived within CFT to take on a universal value, quantified by the central charge $c$ (e.g., $c=1/2$ for the case of the 2D Ising UC)~\cite{Cardy1988}. In particular, a corner with inner angle $\alpha$ along an otherwise straight boundary segment of the spatial domain of a CFT contributes $\frac{c}{24}(\alpha/\pi-\pi/\alpha)$ to the total corner term, resulting from a trace anomaly in the stress tensor.
For a parallelogram-shaped domain the total corner term reads in terms of the complementary inner angles $\alpha$ and $\pi-\alpha$,  
\begin{equation}\label{Eq:fccft}
f_\mathrm{corner}^\mathrm{CFT}(\alpha)=-\frac{c}{12}\left(1+\frac{\alpha}{\pi-\alpha}+\frac{\pi-\alpha}{\alpha}\right),
\end{equation}
which for a rectangle,   $\alpha=\pi/2$, yields the maximum value  of 
$f_\mathrm{corner}^\mathrm{CFT, rec}=-c/4$.
Both expressions are independent of the aspect ratio $\rho=L_2/L_1$ of the edge lengths $L_1$ and $L_2$ of the considered domain. 
Later, Kleban and Vassileva~\cite{Kleban1991} derived a  universal contribution also to $f^c_2$, for conformal invariant critical systems on a rectangle,
$f_2^\mathrm{CFT, rec}(\rho)=-\frac{c}{4}\ln[\eta(\exp(-2\pi\rho)) \eta(\exp(-2\pi/\rho)] + C$,
that depends on $\rho$~\cite{Wu2012}. Here,  $\eta$ is the Dedekind eta function, and  $C$ a number that cannot be computed by CFT methods, but which was later determined for the Ising model using numerical exact solutions~\cite{Wu2012}.
Therefore, the CFT results  for $f^c_\mathrm{corner}$ and $f^{c,\mathrm{rec}}_2$ may be considered universal for conformal invariant 2D systems, apart from the geometric dependence on $\alpha$ and $\rho$,
However, it has been demonstrated that in weakly-anisotropic critical systems various quantities that take on universal values in the isotropic case relevant for CFT, such as critical Binder ratios, Casimir forces or  free energies contributions, can  in fact be strongly affected by the presence of anisotropies in the  critical fluctuations~\cite{Chen2004,Selke2005,Dohm2006,Chen2007,Dohm2008,Selke2009,Kastening2010,Dantchev2009,Dohm2011,Dohm2018,Dohm2019,Dohm2021,Dohm2021a,Sushchyev2023, Dohm2023a, Dohm2023b, Dohm2023c,DANTCHEV20231}. Since spatial anisotropy is omnipresent in condensed matter physics of, e.g., magnetic materials, superconductors, or liquid crystals, it is of fundamental important to account for its effects. 
For example, as predicted in Ref.~\cite{Chen2004}  and demonstrated explicitly below, the value of the corner contribution $f^c_\mathrm{corner}$ in the 
expansion ~(\ref{Eq:fobc}) in general depends on the anisotropy of the critical fluctuations, and takes on the specific value $f_\mathrm{corner}^\mathrm{CFT}$ essentially only in the isotropic limit. This situation begs the question, (i) how the value of $f^c_\mathrm{corner}$ actually depends on the anisotropy of the critical fluctuations, and (ii) whether this dependence can be quantified by a explicit formula in terms of parameters that specify the anisotropy of the critical fluctuations.

Here, we address these questions based on recent advances in the understanding of weakly-anisotropic critical systems.
Namely, it was found  that in the case of {\it periodic} boundary conditions  anisotropy-dependent free energy contributions (more specifically, the critical excess free energy $f^c_\mathrm{ex}=f^c-f^c_{\rm bulk}$) can  be expressed in terms of  nonuniversal parameters that specify the anisotropic fluctuations at criticality, via CFT-based exact formulae that are available for the isotropic limit of the 2D Ising UC~\cite{Dohm2021}. The resulting expressions exhibit complex self-similar structures, reflecting the modular invariance of the torus partition function in the CFT scaling limit. In the following, we extend these recent investigations to finite systems with {\it free} boundary conditions. More specifically, we show how the approach of Ref.~\cite{Dohm2021} can be employed in order to devise an exact formula for the corner term  $f^{c,\mathrm{rec}}_\mathrm{corner}$ for weakly-anisotropic systems on rectangular domains. Furthermore, we use a numerical exact solution of the anisotropic triangular Ising model in order to assess the obtained analytic expressions. 

The remainder of this article is organized as follows: In  Sec.~\ref{Sec:Analytic}, we explain how the exact analytic expression for 
$f^{c,\mathrm{rec}}_\mathrm{corner}$ can be obtained using the approach of Ref.~\cite{Dohm2021}. Then, we provide a detailed comparison to numerical exact results for the anisotropic triangular Ising model in  Sec.~\ref{Sec:Numeric}. Finally, we apply our analytic expression to the case of the anisotropic triangular Potts model in  Sec.~\ref{Sec:Potts}, before we provide a further discussion and outlook in Sec.~\ref{Sec:Conclusions}.


\section{Analytic expressions for weakly anisotropic systems}\label{Sec:Analytic}

Since our analytical results build upon an approach presented in Ref.~\cite{Dohm2021}, we first summarize the relevant steps. 
It was proposed in Ref.~\cite{Dohm2021}, and later confirmed by Refs.~\cite{Dohm2021a, Dohm2023a, Dohm2023b, Dohm2023c}, that for weakly-anisotropic systems in the 2D Ising UC on a rectangular domain with periodic boundary conditions, the amplitude $\mathcal{F}_c$ in the leading finite-size scaling form $f^c_\mathrm{ex}=\mathcal{F}_c/A+ O(1/A^2)$ of the critical excess free energy can be calculated from the CFT expression for the partition function of the isotropic Ising model on appropriately constructed torus geometries. 
\begin{figure}[t]
    \centering
    \includegraphics[width=0.35\textwidth]{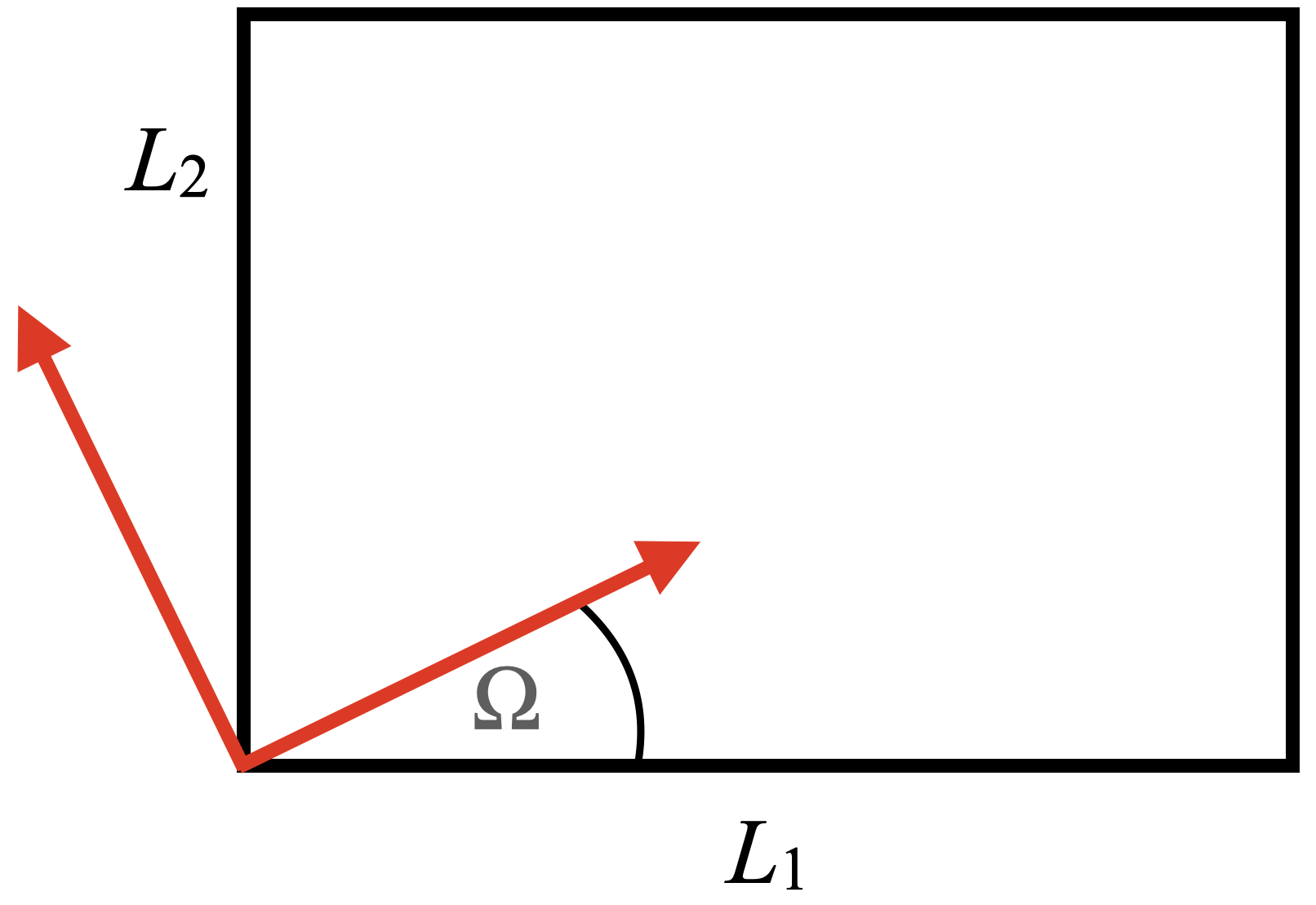}
    \caption{Rectangular domain with edge lengths $L_1$ and $L_2$. The red arrows indicate the two orthogonal principle directions of the bulk critical order-parameter correlation function, specified by the angle $\Omega$.   }
    \label{Fig:rectangle}
\end{figure}
This construction depends on two bulk quantities that specify the anisotropic fluctuations in terms of the order-parameter correlation function in the scaling regime: 
More specifically, for a weakly-anisotropic 2D system, the angular dependence of the critical correlations of the bulk system is given by (i) the angle $\Omega$, specifying  the orientation of the two principal directions, and (ii) the ratio $q$ of the two principal correlation lengths upon approaching criticality~\cite{Dohm2019, Dohm2021}. See Fig.~\ref{Fig:rectangle} for an illustration. 
By use of an effective shear transformation, the original rectangular domain  with aspect ratio $\rho$ is then  mapped onto a parallelogram with angle $\alpha$ and aspect ratio $\rho_\mathrm{p}$, such that under this mapping the correlation function becomes isotropic, making CFT applicable~\cite{Dohm2021, Dohm2023a, Dohm2023b}. The corresponding parallelogram parameters are given by 
\begin{eqnarray}
\label{Eq:alpha}
&& \cot \alpha(q,\Omega)
= (q^{-1}-q)\cos \Omega \;\sin \Omega,
\\
\label{Eq:rhop}
&& [\rho_{\rm{p}}(\rho,q,\Omega)]^2= \rho^2\; \frac{\tan^2\Omega+q^2}{1+q^2\tan^2\Omega}.
\end{eqnarray}
A parallelogram with periodic boundary conditions is topologically equivalent to a torus whose  shape dependence can be parameterized in terms of the complex torus modular parameter $\tau$, with
%
$\tau(\alpha,\rho_{\rm{p}})=\rho_{\rm{p}}\exp(i\; \alpha)$,
%
and from CFT~\cite{DiFrancesco1988,DiFrancesco1997} it is known that the critical amplitude of the free energy on the torus is exactly given by
%
${\cal F}_c^{\rm{CFT}}(\tau)= -\ln Z^{\rm{CFT}}(\tau)$
%
in terms of the partition function $Z^{\rm{CFT}}(\tau)$ of the isotropic 2D Ising model. The latter can in fact be expressed as
$Z^{\rm{CFT}}(\tau)=\big({|\theta_2(\tau)|+|\theta_3(\tau)|+|\theta_4(\tau)|}\big)/\big({2|\eta(\tau)|}\big)$
in terms of Jacobi theta functions $\theta_i(0|\tau)\equiv \theta_i(\tau)$ and the Dedekind eta function. 
In summary, these steps lead to the explicit formula
$\mathcal{F}_c(\rho, q,\Omega)=-\ln Z^{\rm CFT}(\tau(\alpha(q,\Omega),\rho_\mathrm{p}(\rho,q,\Omega)))$
of Ref.~\cite{Dohm2021} for the critical amplitude of the excess free energy for  anisotropic models in the 2D Ising UC. 

\begin{figure}[t]
    \centering
    \includegraphics[width=0.45\textwidth]{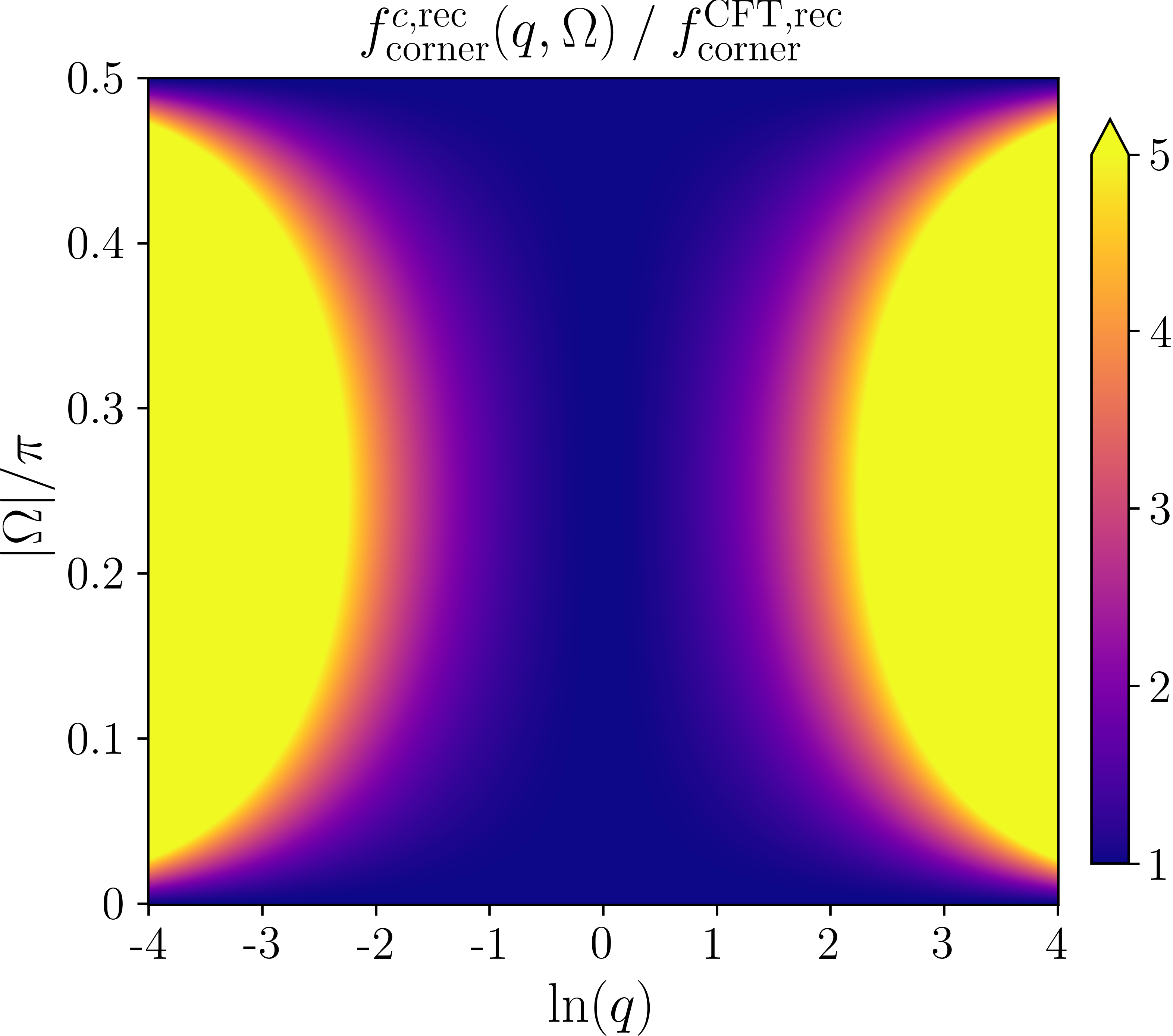}
    \caption{Dependence of the corner free energy term $f_\mathrm{corner}^{c,\mathrm{rec}}$ in units of the CFT result $f_\mathrm{corner}^\mathrm{CFT,rec}=-c/4$ on the anisotropy parameters $q$ and $\Omega$ for a rectangular domain from Eq.~(\ref{Eq:fcrec}).}
    \label{Fig:fcqOmega}
\end{figure}

Returning to the case of free boundary conditions, we can employ the same shear transformation in order to relate the original, weakly-anisotropic model on the rectangular domain to an isotropic model on the parallelogram, which is again specified by Eqs.~(\ref{Eq:alpha}) and ~(\ref{Eq:rhop}). We  then use Eq.~(\ref{Eq:fccft}) to obtain the corner contribution for the anisotropic model on the rectangular domain in terms of the CFT result on the parallelogram, such that
\begin{equation}\label{Eq:fcrec}
f_\mathrm{corner}^{c,\mathrm{rec}}(q,\Omega)=f_\mathrm{corner}^\mathrm{CFT}(\alpha(q,\Omega)).
\end{equation}
This formula is the central result of this work and in the next section we will compare it to numerical exact data for a specific anisotropic Ising model. 
Before doing so, we illustrate in Fig.~\ref{Fig:fcqOmega} the dependence of the geometric ratio $f_\mathrm{corner}^{c,\mathrm{rec}}/f_\mathrm{corner}^\mathrm{CFT,rec}$ on the anisotropy parameters $q$ and $\Omega$ that follows from Eq.~(\ref{Eq:fcrec}). Like the CFT result, this formula does not depend on the aspect ratio $\rho$ of the rectangle.
 The figure directly illustrates the nonuniversal character of the corner term for weakly-anisotropic critical 2D systems. From Eq.~(\ref{Eq:fcrec}), the CFT result $f_\mathrm{corner}^\mathrm{CFT,rec}=-c/4$ is recovered in the isotropic limit $q=1$, as well as for $\Omega=0$ and $\pm\pi/2$, i.e., when the principal axes of the correlation ellipsoid align parallel to the edges of the rectangular domain (irrespective of the value of $q$). 
 In general the nonuniversal quantities $q$ and $\Omega$ exhibit a non-trivial  dependence on microscopic details, e.g., the couplings in the case of an Ising model (an explicit example will be presented in the following section). However, expressed in terms of $\alpha(q,\Omega)$, the above formula instead  provides a universal relation for the corner contribution via the central charge of the underlying CFT scaling limit. We note that in contrast to the case of $f^c_\mathrm{ex}$ for periodic boundary conditions~\cite{Dohm2021}, no self-similar structures appear in  Fig.~\ref{Fig:fcqOmega}, as expected form the absence of modular invariance for free boundary conditions. 

\section{Comparison to numerical exact results}\label{Sec:Numeric}

\begin{figure}[t]
    \centering
   \includegraphics[width=0.4\textwidth]{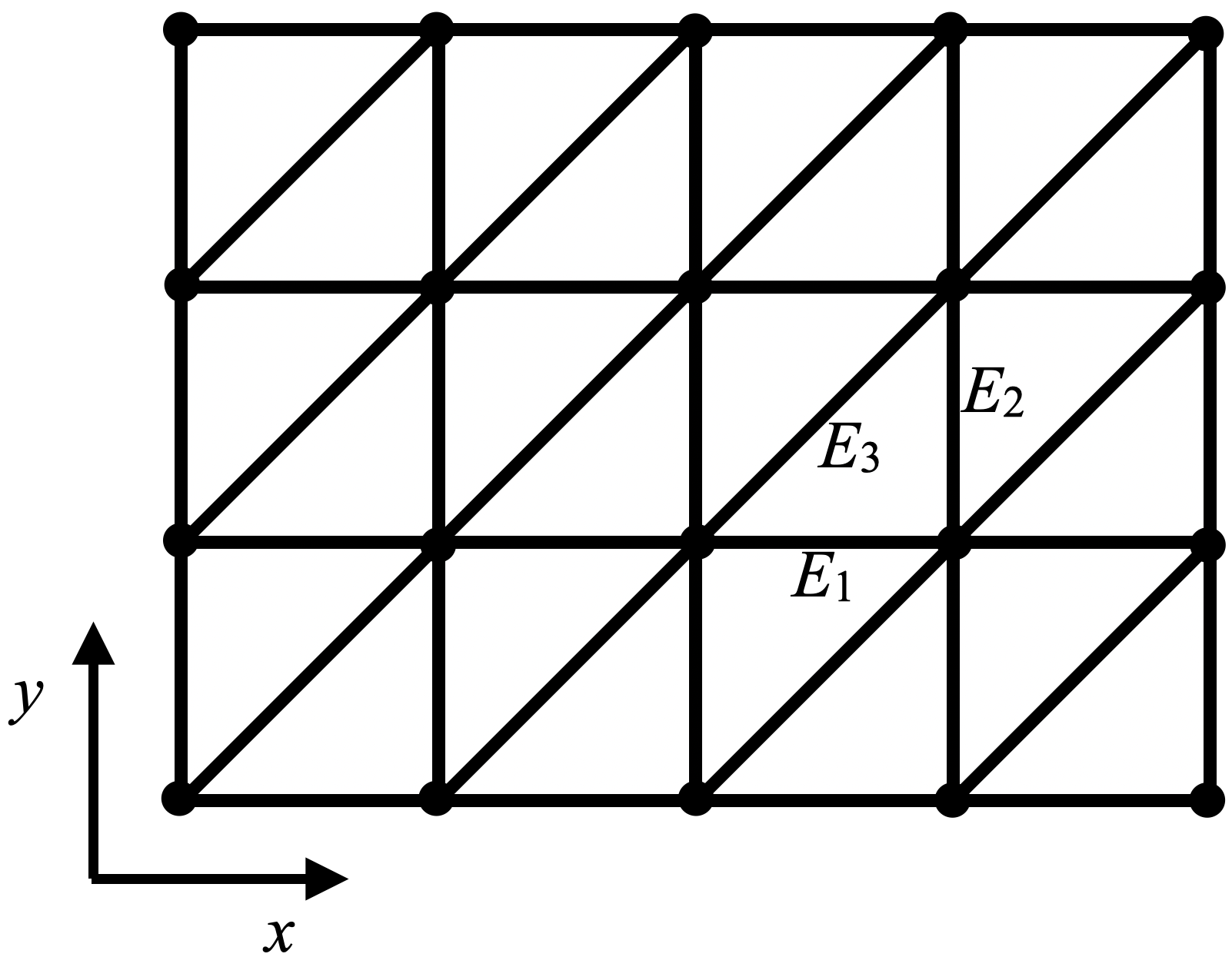}
   \caption{Illustration of the  anisotropic triangular Ising model with couplings $E_1$, $E_2$ and $E_3$ on a $L_1\times L_2= 5\times 4$ square lattice (lattice constant $a_0=1$).}
   \label{Fig:lattice}
\end{figure}

For this purpose, we consider the anisotropic triangular Ising model~\cite{Stephenson2004, Dohm2019, Dohm2021, Dohm2021a}, defined by
\begin{equation}
\label{IsingH}
H\! =\!-\!\!\sum_{i}\big [E_1\sigma_{i} \sigma_{i+\hat{x}}+E_2\sigma_{i} \sigma_{i+\hat{y}}  +E_3\sigma_{i} \sigma_{i+\hat{x}+\hat{y}}\big],
\end{equation}
where the spin variables $\sigma_{i}=\pm1$ reside on a square lattice with horizontal, vertical, and (up-right) diagonal couplings $E_1, E_2$, $E_3$, cf. Fig.~\ref{Fig:lattice} for an illustration. 
In particular, we consider the ferromagnetic regime of $H$, where  the system exhibits a thermal phase transition to a low-temperature ferromagnetic phase in the thermodynamic limit. This regime is constrained by three simultaneous conditions: $E_1+E_2>0$, $E_1+E_3>0$, and $E_2+E_3>0$ on the three couplings $E_1,E_2,E_3$.
The condition for the critical temperature $T_c$, separating the low-$T$ ferromagnetic phase from the paramagnetic regime, reads 
$\hat S_1 \hat S_2+\hat S_2 \hat S_3+\hat S_3 \hat S_1=1$, where
$\hat S_\alpha= \sinh (2 E_\alpha/k_BT_c)$~\cite{Houtappel1950}.
In order to extract the corner contribution to the free energy, we considered this model on finite rectangular lattices with $N=L_1 \times L_2$ lattice sites, fixing the lattice constant to $a_0=1$, such that $A=L_1 L_2$, and $E=2(L_1+L_2)$. The corresponding finite lattices with free (open) boundary conditions are illustrated in Fig.~\ref{Fig:lattice}. In the following, we focus on $\rho=1$. 

We obtain numerical exact values of the critical free energy density
\begin{equation}
f^c= \frac{1}{A} \ln Z_c, \quad Z_c=\sum_{\sigma_1,...,\sigma_N} \exp(-\beta_c H),
\end{equation}
where $\beta_c=1/(k_BT_c)$, based on the Grassmann variable approach used by Plechko~\cite{Plechko1985,Plechko1988,Plechko1996}.
Alternatively, one can use  the bond-propagation algorithm~\cite{Loh2006} for this purpose. 
We then extract the corner term and other finite-size as well as the bulk contribution from fitting, for fixed couplings and aspect ratio, the finite-size data for different system sizes to the expansion in Eq.~(\ref{Eq:fobc}), which we reproduce here, now including also higher order terms, 
\begin{equation}\label{Eq:fobcfull}
f^c=f^c_{\rm bulk} + f^c_{\rm surface} \frac{1}{E} + f^c_\mathrm{corner} \frac{\ln A}{A} + \sum_{k=2}^\infty f^c_k \frac{1}{A^{k/2}}.
\end{equation}
In practice we truncate the series at a maximum order of $k\leq k_{\max}=8$ and use the Levenberg-Marquardt method for performing the corresponding non-linear fitting.   

\begin{figure}[t]
    \centering
 \includegraphics[width=0.45\textwidth]{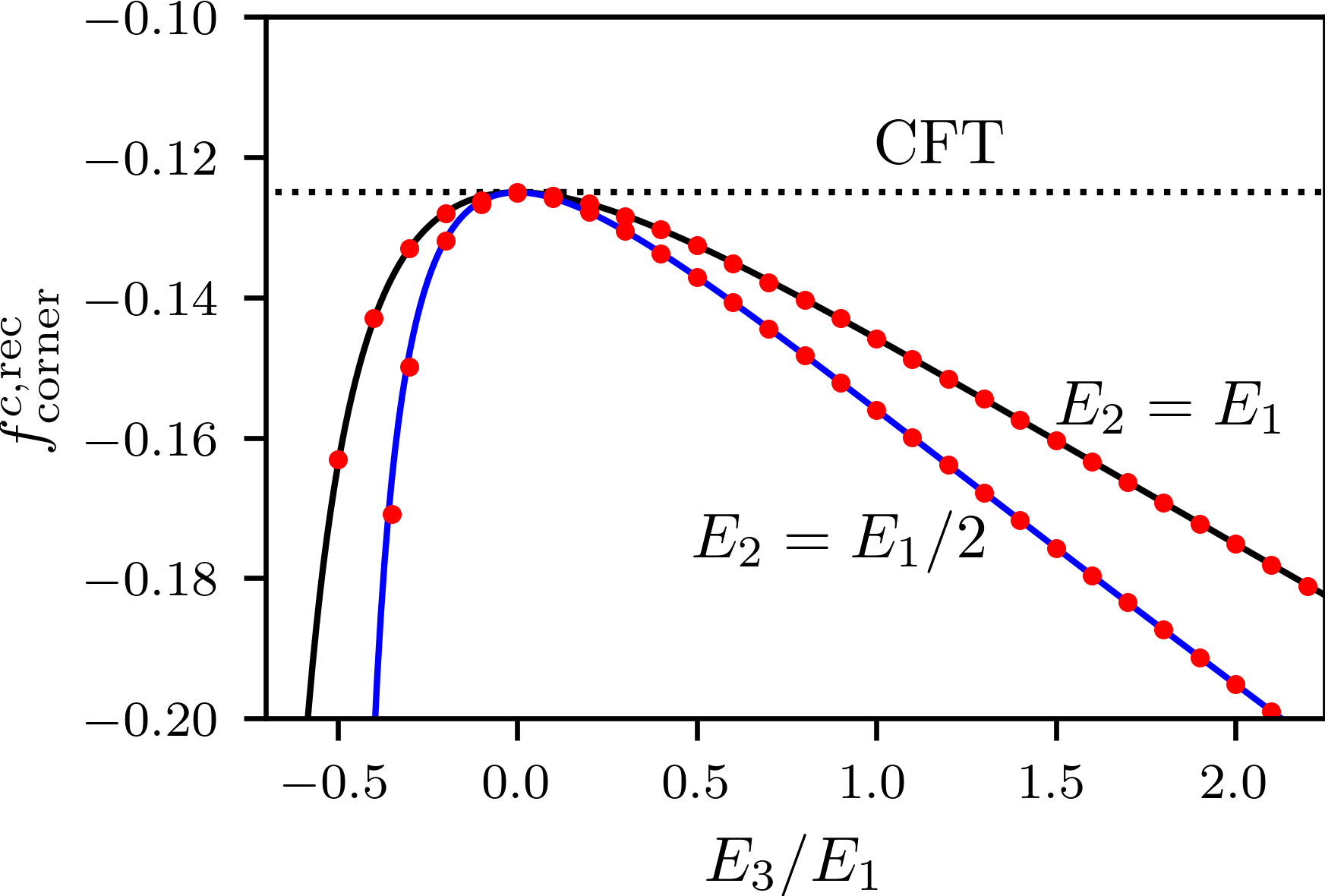}
   \caption{Comparison of the numerical estimates (circles) for $f^{c,\mathrm{rec}}_\mathrm{corner}$ with the CFT-based formula (solid lines) in Eq.~(\ref{Eq:fcrec}) for the anisotropic triangular Ising model upon varying $E_3$ along the two lines  $E_2=E_1$ and $E_2=E_1/2$, based on numerical exact solutions for system sizes up to $70 \times 70$. The dashed line indicates the CFT value of $-1/8$ for the 2D Ising UC.}
   \label{Fig:compare} 
\end{figure}

\begin{figure}[t]
    \centering
    \includegraphics[width=0.45\textwidth]{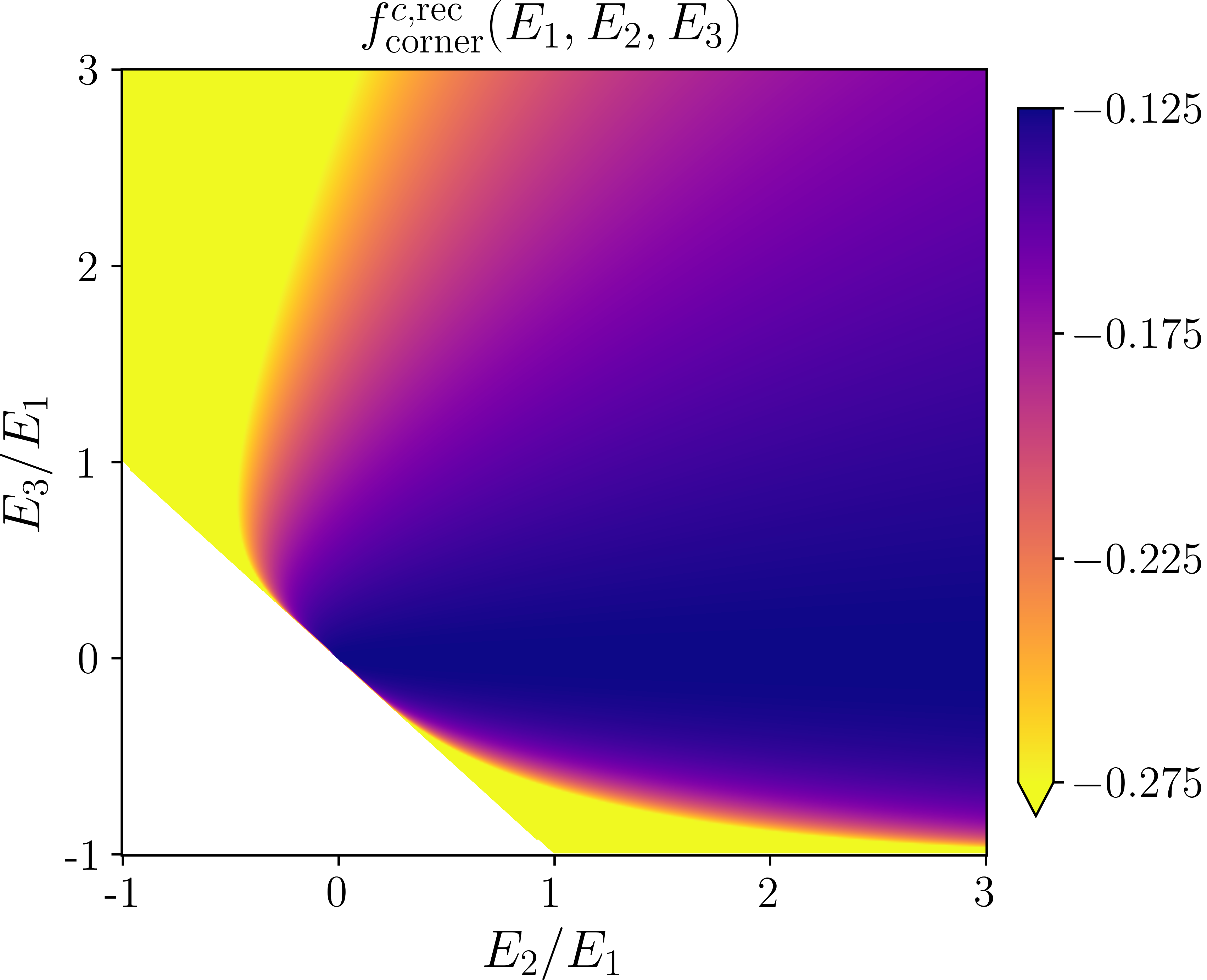}
   \caption{Corner contribution $f^{c,\mathrm{rec}}_\mathrm{corner}$ based on Eq.~(\ref{Eq:fcrecIs}) as a function of the couplings inside the ferromagnetic regime (outside the white area) of the anisotropic triangular Ising model.}
   \label{Fig:fcIsing}
\end{figure}

In order to compare the numerical estimates for $f^{c,\mathrm{rec}}_\mathrm{corner}$ to the CFT-based prediction from Eq.~(\ref{Eq:fcrec}), explicit values of $q$ and $\Omega$ for the anisotropic Ising model described by $H$ are required. Closed formulae for both quantities have  been obtained recently~\cite{Dohm2019} and read 
\begin{eqnarray}
    \tan(2\Omega)&=&\frac{2(1-\hat S_1 \hat S_2) }{\hat S_2^2 -\hat S_2^2}, \quad {\rm for }\:\:  E_1\neq E_2,\\
    \Omega&=&\pi/4, \quad {\rm for }\:\:  E_1 = E_2,
\end{eqnarray}
and, for $ E_1\neq E_2$, 
\begin{equation}
    q=\frac{2+\hat S_1^2+\hat S_2^2\pm [(\hat S_1^2+\hat S_2^2)^2+4(1-2\hat S_1 \hat S_2)]^{1/2}}{2(\hat S_1+\hat S_2)},
\end{equation}
where the sign in front of the square root depends on whether $E_1>E_2$ $(+)$ or $E_1<E_2$ $(-)$,
while 
\begin{equation}
q=1/\hat S_1, \quad {\rm for }\:\:  E_1 = E_2,
\end{equation}
respectively. Inserting these expressions into Eq.~(\ref{Eq:fcrec}), we obtain a compact result
\begin{equation}\label{Eq:fcrecIs}
f_\mathrm{corner}^{c,\mathrm{rec}}(E_1,E_2,E_3)=f_\mathrm{corner}^\mathrm{CFT}(\mathrm{arccot}(\hat S_3))
\end{equation}
for the anisotropic triangular Ising model. Upon tuning the coupling ratios, one can realize any possible value of the corner term for the 2D Ising UC in this model. 
A comparison of the numerical estimates for $f^{c,\mathrm{rec}}_\mathrm{corner}$ to this CFT-based formula for the two cases of $E_2=E_1$ and $E_2=E_1/2$ and varying $E_3$ is provided in Fig.~\ref{Fig:compare} (in both cases the ferromagnetic regime is restricted to  $E_3>-E_2$). We find excellent agreement between the numerical results and our analytic expression. This is observed at other parameter values and aspect ratios as well, though the numerical values become less accurate for negative values of $E_3$, where frustration emerges and the correlations become increasingly anisotropic upon approaching the limit of weak anisotropy. In Fig.~\ref{Fig:compare} the CFT value,
which equals  $-1/8$ for $c=1/2$, is recovered only in the specific case of $E_3=0$. At this point, isotropy is restored in the scaling limit for $E_2=E_1$, i.e., $q=1$. For $E_2=E_1/2$, the critical fluctuations are instead anisotropic also for $E_3=0$, with $q=1.543...$. However, at this point the principle axes align parallel to the lattice directions, $\Omega=0$, thus leading to the CFT value.   
Finally, Fig~\ref{Fig:fcIsing} shows the microscopic parameter dependence of $f^{c,\mathrm{rec}}_\mathrm{corner}$ according to Eq.~\ref{Eq:fcrecIs} within  the ferromagnetic region of the anisotropic triangular Ising model at criticality. Here, we again observe a non-trivial dependence on the microscopic parameters, as anticipated in the previous section. In particular, upon approaching the boundary of the ferromagnetic domain, where weak anisotropy breaks down, we find increasingly large deviations from the CFT result, see also Fig.~\ref{Fig:compare}. Indeed, it follows from Eq.~\ref{Eq:fcrecIs} that  
$f^{c,\mathrm{rec}}_\mathrm{corner}\rightarrow -\infty$ in this regime. 

\section{Anisotropic Triangular Potts Model}\label{Sec:Potts}
\begin{figure}[t]
    \centering
    \includegraphics[width=0.45\textwidth]{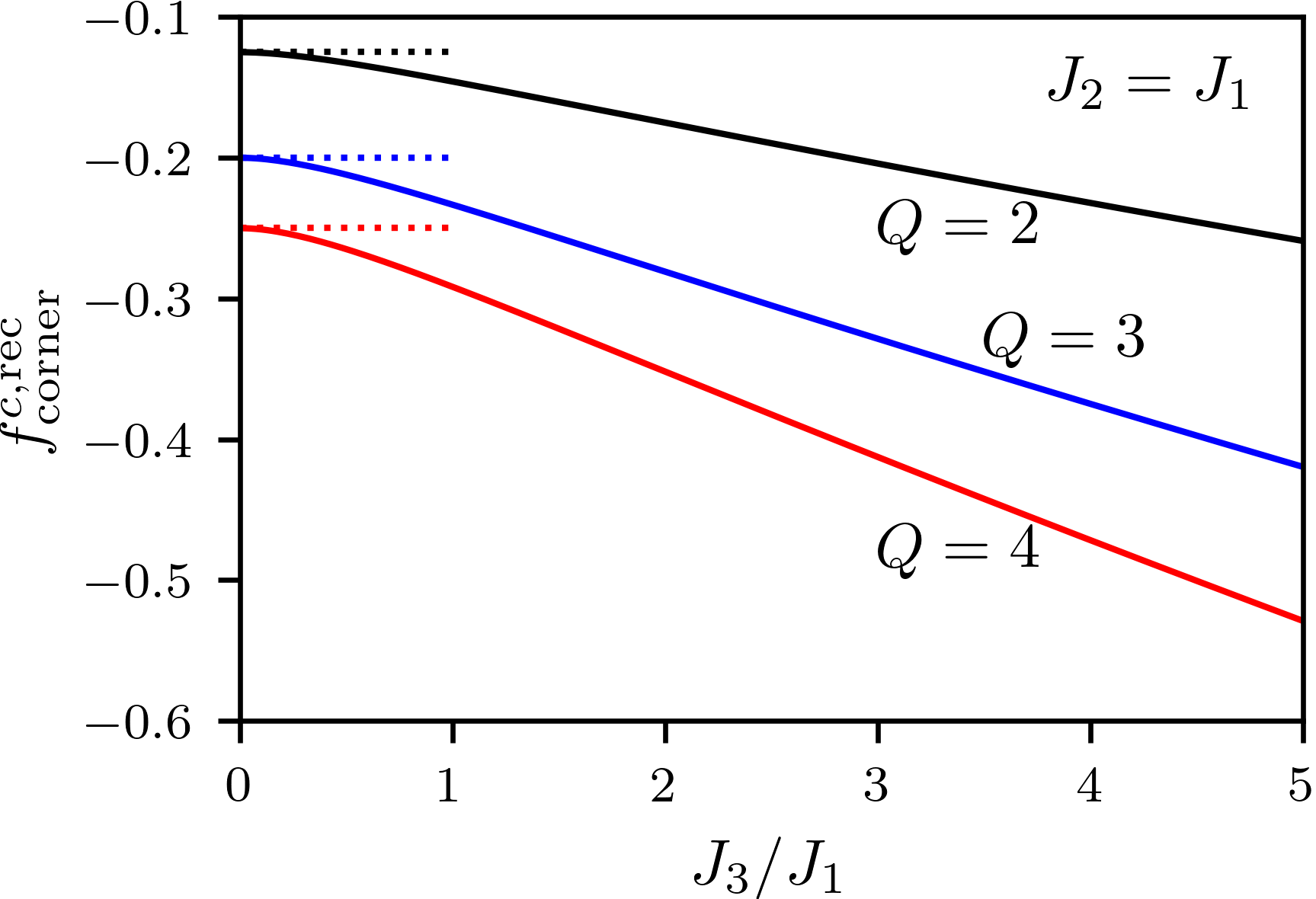}
   \caption{Corner contribution $f^{c,\mathrm{rec}}_\mathrm{corner}$ for the anisotropic $Q$-states Potts model as a function of $J_3$ along the line $J_2=J_1$ for $Q=2,3$ and $4$. Dashed lines denote the corresponding CFT values $-c/4$, where $c=1/2,4/5,1$ for $Q=2,3,4$, respectively.}
   \label{Fig:Potts}
\end{figure}

As a further application of our formula for the corner term,  Eq.~(\ref{Eq:fcrec}), we consider the case of the $Q$-states Potts model~\cite{Wu1982} on the  rectangular lattice of Fig.~\ref{Fig:lattice}, with the Hamiltonian
\begin{equation}
\label{PottsH}
H_Q\! =\!-\!\!\sum_{i}\big [J_1\:\delta_{n_i,n_{i+\hat{x}}}+J_2\:\delta_{n_i,n_{i+\hat{y}}}
 +J_3\:\delta_{n_i,n_{i+\hat{x}+\hat{y}}}\big],
\end{equation}
where $n_i=1,...,Q$ and $\delta$ denotes the Kronecker symbol, and $J_i\geq 0$, $i=1,2,3$. While the case $Q=2$ corresponds to the triangular Ising model discussed above, the Potts model exhibits a second-order thermal phase transition also for $Q=3$ and $4$. 
The latter are described by CFTs with a central charge $c$ of $4/5$ and $1$, respectively. While the condition 
$\sqrt{Q}\: x_1 x_2 x_3 + x_1 x_2 + x_2 x_3 + x_3 x_1 =1$, where $x_i=(e^{2\beta_c J_i}-1)/\sqrt{Q}$, for the critical inverse temperature $\beta_c$ is well known (though not proven for $Q=3$)~\cite{Wu1982}, only recently were exact expressions reported that allow us to determine the parameters $\rho_\mathrm{p}$ and $\alpha$, specifying the shear transformation for the general anisotropic case. More specifically, in Sec.~V of their supplemental material, the authors of Ref.~\cite{Hu2022} consider the triangular Potts model on $L\times L$ lattices and specify shear transformations in terms of an effective aspect ratio $\rho_{\rm{e}}$ and a boundary twist $t_{\rm{e}}$, using the isoradial-graph method~\cite{Hugo2018}. Both quantities can be expressed in terms of $\rho_\mathrm{p}$ and $\alpha$, $\rho_{\rm{e}}=\rho_{\rm{p}}\sin(\alpha)$, $t_{\rm{e}}=\rho_{\rm{p}}\cos(\alpha)$, as we verified for the case $Q=2$ based on the results of Refs.~\cite{Dohm2019, Dohm2021}. Using this correspondence for general $Q$, we obtain the following relation for $\alpha$: 
\begin{equation}
\label{Pottsalpa}
e^{2\beta_c J_3}-1 =\begin{cases} \sqrt{Q} \frac{\sin(r\: (\pi- 2 \alpha))}{\sin( 2 \: r\: \alpha)}, & Q=2,3,\\ \frac{\pi- 2 \alpha}{\alpha}, & Q=4, 
\end{cases}
\end{equation}
where $r=\arccos(\sqrt{Q}/2)/\pi$.
Determining $\alpha$ from (numerically) solving this equation and inserted into Eq.~(\ref{Eq:fcrec}) yields the corner contribution for the anisotropic Potts model on a rectangular domain, shown for $Q=2,3$ and $4$ in Fig.~\ref{Fig:Potts}, varying $J_3$ along the line $J_2=J_1$ (the results for $Q=2$ agree with those in Fig.~\ref{Fig:compare}). In all cases we observe a similarly strong suppression of $f_\mathrm{corner}^{c,\mathrm{rec}}$ for finite values of $J_3$ from the CFT value $-c/4$, which is recovered only for $J_3=0$. 

\section{Conclusions}\label{Sec:Conclusions}
Using the effective shear transformation from Refs.~\cite{Dohm2021, Dohm2023a, Dohm2023b} to map a weakly-anisotropic critical 2D system onto an isotropic one, we obtained an extension of the CFT prediction~\cite{Cardy1988} for the corner contribution to the critical free energy on rectangular domains. This formula, Eq.~(\ref{Eq:fcrec}), explicitly demonstrates the nonuniversal character of the corner term for anisotropic systems, as its value depends strongly on the parameters $q$ and $\Omega$ that characterize the anisotropic critical fluctuations, and which need to be calculated for each specific microscopic model. We find that the CFT value is recovered only for the isotropic case ($q=1$), or when the principle axes align with the edges of the rectangular domain ($\Omega=0, \pi/2$).   Moreover, the resulting expression was found to be in accord with numerical exact results obtained for the anisotropic triangular Ising model. We expect this formula to apply also to other weakly-anisotropic critical systems that allow for a CFT description in the isotropic limit via  appropriate shear transformations, and considered the triangular Potts model as a futher example.  

These findings suggest several directions for further investigations: While we focused here on rectangular domains, it is also feasible to generalize our approach to the case of anisotropic systems on a parallelogram domain, using a corresponding shear transformation to another, isotropic parallelogram~\cite{Dohm2023c}. In addition, it would also be important to extend these investigations towards exploring corner contributions beyond the critical point, similar to the analysis performed for periodic boundary conditions in Ref.~\cite{Dohm2023c}. Furthermore, it would be interesting to explore  the anticipated universal contribution to $f^{c,\mathrm{rec}}_2$ proposed in Ref.~\cite{Kleban1991} from CFT towards the anisotropic case. This could indeed be realized based on the methodology that was used here, but then requires a generalization from $f^{\mathrm{CFT,rec}}_2$ to the case of CFTs on general parallelograms, i.e., beyond the rectangular domains considered in Ref.~\cite{Kleban1991}. We are not aware of such a generalization and leave all the above directions for future research. 

\section*{Acknowledgements}
We thank Volker Dohm for collaborations on related projects. Furthermore,  we acknowledge support by the Deutsche Forschungsgemeinschaft (DFG) through RTG 1995, and thank the IT Center at RWTH Aachen University for access to computing time.

\bibliography{main.bbl}
\end{document}